\documentclass[a4paper,aps,pre,floatfix,nofootinbib,twocolumn,superscriptaddress,showpacs]{revtex4-1}

\usepackage{graphicx}
\usepackage{latexsym,amsmath,verbatim}
\usepackage{color}
\usepackage{hyperref}

\newcommand{\av}[1]{\langle {#1} \rangle}

\newcommand{\eps}{\varepsilon}

\begin{document} 

\title{Universal and non-universal features of the generalized voter
  class for ordering dynamics in two dimensions}

\author{Claudio Castellano}

\affiliation{Istituto dei Sistemi Complessi (ISC-CNR), via dei Taurini
  19, I-00185 Roma, Italy}

\affiliation{Dipartimento di Fisica, ``Sapienza'' Universit\`a di
  Roma, P.le A. Moro 2, I-00185 Roma, Italy}

\author{Romualdo Pastor-Satorras}

\affiliation{Departament de F\'\i sica i Enginyeria Nuclear,
  Universitat Polit\`ecnica
 de Catalunya, Campus Nord B4, 08034
  Barcelona, Spain}

\date{\today}

\begin{abstract}
  By considering three different spin models belonging to the
  generalized voter class for ordering dynamics in two dimensions
  [I. Dornic, \textit{et al.}  Phys. Rev. Lett. \textbf{87}, 045701
  (2001)], we show that they behave differently from the linear voter
  model when the initial configuration is an unbalanced mixture up and
  down spins.  In particular we show that for nonlinear voter models
  the exit probability (probability to end with all spins up when
  starting with an initial fraction $x$ of them) assumes a nontrivial
  shape. This is the first time a nontrivial exit probability is
  observed in two dimensional systems.
  The change is traced back to the strong nonconservation of
  the average magnetization during the early stages of dynamics.  Also
  the time needed to reach the final consensus state $T_N(x)$ has an
  anomalous nonuniversal dependence on $x$.
\end{abstract}

\pacs{05.40.-a, 89.65.-s, 64.60.De}

\maketitle

\section{Introduction}
\label{sec:introduction}

The voter model (VM) \cite{Clifford73,Holley:1975fk} is a paradigm of
coarsening phenomena \cite{Bray94} that stands as one of the most
interesting models in non-equilibrium statistical mechanics
\cite{KineticViewRedner}. The nature of its interest is twofold: On
the one hand, it represents one of the few nontrivial non-equilibrium
statistical processes that can be exactly solved in any number of
dimensions \cite{liggett99:_stoch_inter,PhysRevE.53.R3009}.  On the
other hand, it has a natural application in social dynamics
\cite{Castellano09} as a model for the formation of opinion consensus
in a society initially divided in two different standpoints.  The
appeal of the VM is further enhanced by its connection with neutral
models in genetics, ecology and
linguistics~\cite{Crowbook,hubbell2001unified,Blythe09}.  Its
definition is very simple: On a regular lattice or graph, each site is
endowed with a binary variable $s_i=\pm 1$. At each time step, a
randomly chosen site copies the state of one of its nearest neighbors,
chosen in its turn at random. This parameter-free dynamics can be
  succinctly encoded in the \textit{flipping probability} $f(x_i)$,
measuring the probability that spin $i$ will flip if surrounded by a
fraction $x_i$ of spins in the opposite state, which takes the simple
linear form $f(x_i)=x_i$.  For this reason we will refer to it in the
following as the {\em linear} voter model. Voter dynamics is thus
characterized by the presence of two absorbing states (all spins
either $+1$ or $-1$, the \textit{consensus} states) with a $Z_2$ spin
reversal symmetry. Moreover, since the rate of creation of $+1$ and
$-1$ spins is equal, the magnetization is conserved in average.

The way in which consensus is reached in the VM can be characterized
from different perspectives.  From the point of view of
non-equilibrium statistical mechanics, the coarsening process in the
VM is marked by the absence of surface tension
\cite{PhysRevLett.87.045701} causing an anomalous logarithmic decay of
the density of interfaces in $d=2$, namely $\rho(t) \sim 1/\ln(t)$, in
opposition to curvature-driven dynamics \cite{Bray94}, which leads to
an algebraic decay $\rho(t) \sim t^{-1/2}$.  In the social dynamics
context, on the other hand, interest is focused on the exit
probability $E(x)$~\cite{Castellano09} (defined as the probability
that the final state corresponds to all sites in state $+1$) and the
consensus time $T_N(x)$ (the average time needed to reach consensus in
a system of size $N$) when starting from fully random initial
conditions with a fraction $x$ of sites in state $+1$. Conservation of
magnetization implies a characteristic linear exit probability, $E(x)
=x$, in any dimension $d$ \cite{KineticViewRedner}, while the
consensus time takes the form, for $d>1$,
\cite{1751-8121-43-38-385003}
\begin{equation}
  \label{eq:2}
  T_{N}(x) = - N_\mathrm{eff} [x \ln(x) + (1-x) \ln(1-x)],  
\end{equation}
where $N_\mathrm{eff}$ is an effective factor depending on the number
of sites $N$ and dimensionality $d$ \cite{KineticViewRedner}.

A detailed analysis revealed that VM actually lies at the transition
point between a ferromagnetic (ordered) and a paramagnetic
(disordered) phase, such that infinitesimally small perturbations are
able to drastically change its behavior
\cite{Deoliveira93,Drouffe99,Molofsky99}. The parameter-free nature of
the VM thus led naturally to the question as whether it represents a
peculiar and isolated point, or rather belongs to a more general
(universal) class of models, sharing the same properties. This issue
has been answered by Dornic and coworkers
\cite{PhysRevLett.87.045701,AlHammal05} (see
also~\cite{Vazquez08,PhysRevLett.95.100601}), who have pointed out the
existence of a genuine generalized voter (GV) universality class,
encompassing systems at an order-disorder transition driven only by
interfacial noise, between two ``dynamically equivalent'' absorbing
states. The dynamical equivalence between states can be enforced
either by $Z_2$-symmetric local rules, or by global conservation of
the magnetization.  The linear voter model possesses both
properties, but each of them separately is sufficient to ensure GV
behavior.  The GV class is characterized in $d=2$ by the logarithmic
decay of the density of interfaces\footnote{The behavior of the VM in
  $d=1$ coincides with the zero temperature Glauber dynamics
  \cite{KineticViewRedner}, while it is described by mean-field theory
  for $d>2$, its upper critical dimension. The interest of its
  definition and properties is thus essentially given by the behavior
  in $d=2$.}  as well as by other critical
exponents~\cite{PhysRevLett.87.045701}.  In this generalized
perspective, the GV transition for $Z_2$-symmetric models can be
theoretically rationalized as the superposition of two independent
transitions \cite{Drofeli,AlHammal05}, an Ising transition and a
directed percolation \cite{marro1999npt} transition, whose respective
symmetries are broken in unison at the GV manifold.  These
$Z_2$-symmetric models are also called {\em nonlinear} voter models,
because at the transition the flipping probability $f(x_i)$ assumes a
nonlinear form.

By means of extensive numerical simulations performed for three
  representative models, in this paper we show that, while the GV
  class is well-defined in two dimensions in terms of the decay of the
  density of interfaces and the value of a set of the critical
  exponents, it also exhibits non-universal properties which depend on
  the microscopic details of the respective models' definitions.  The
  non-universality of the GV class is explicitly observed in the exit
  probability and the consensus time, which deviate from the linear
  and entropic form [Eq.~\eqref{eq:2}], respectively, observed in the
  linear voter model. In this respect, it is worth noticing that
  nontrivial shapes of the exit probability had previously been found
  for models in $d=1$ or at the mean-field level. 
  Here we show for the first time that $E(x)$ can also be
  nontrivial in two-dimensional systems.

  We have considered in particular three models representing the whole
  spectrum of GV class, namely the nonlinear voter model (NLV)
  originally devised to explore the GV manifold
  \cite{PhysRevLett.87.045701}; the recently proposed non-linear
  $q$-voter model (qV) \cite{PhysRevE.80.041129}; and the Kaya,
  Kabak{\c c}io{\v g}lu, and Erzan (KKE) model~\cite{KKEmodel00}.  The
  first two models are $Z_2$-symmetric, while in the third the
  dynamical equivalence between absorbing states is enforced by global
  conservation of magnetization. Numerical simulations in the vicinity
  of the critical point confirm the existence of the GV universality
  class. In particular, measuring the exponents related to the
  fluctuations of the magnetization and the correlation length when
  approaching the critical point from the disordered and ordered
  phases, respectively, suggest non mean-field exponents, at odds
  with the claim made in
  Ref.~\cite{PhysRevLett.87.045701}. However, when probing the
  behavior of the models for unbalanced initial conditions ($x \neq
  1/2$) by means of the dependence of the exit probability and of the
  consensus time on $x$, we observe that the originally defined GV
  class exhibits strong non-universal features, represented by an exit
  probability and the consensus time that can depend on further
  microscopic details of the models undergoing the GV transition. In
  particular, we observe that models in which conservation of
  average magnetization is strictly enforced, such as the KKE model,
  have indeed a linear $E(x)$ as the VM, but they exhibit a consensus
  time $T_N(x)$ different from the entropic form Eq.~\eqref{eq:2}. On
  the other hand, models which exhibit $Z_2$-symmetry, such as the NLV
  and the qV models, display $E(x)$ and $T_N(x)$ both departing from
  the linear VM behavior.  The non-linearity of the exit probability can be
  rationalized by inspecting the behavior of the average magnetization
  over time. Here we can see that magnetization is not conserved over
  short time scales, but it increases initially in the NLV model,
  while it decreases in the qV model. This transient behavior can be
  traced back to the presence of strong non-zero drift in the initial
  dynamical evolution starting from $x\neq 1/2$. After this initial drift
  has vanished, the average magnetization remains constant and the ensuing
  evolution is well described by a linear voter dynamics.
  
The paper is organized as follows: In
Section~\ref{sec:gener-voter-class} we present the results of
numerical simulations for the class of non-linear voter models,
determining the values of the critical exponents and showing the
nontrivial $x$ dependence of the exit probability and of the consensus
time.  In Section~\ref{sec:q-voter-model} we do the same for the
$q$-voter model, while Section~\ref{sec:kaya-kabakc-ciov} is devoted
to the KKE model.  The final Section summarizes the results and
discusses their relevance.
  
\section{The Non-linear Voter model}
\label{sec:gener-voter-class}

The characteristics of the GV class were exposed in
Ref.~\cite{PhysRevLett.87.045701} by numerical examination of a
$Z_2$-symmetric non-linear voter model (NLV) defined it terms of a
kinetic Ising model as follows: We consider a binary spin system in
$d=2$, in which the probability $r_{s,h}$ that a spin $s$ flips
depends on its value and the value of the local field $h$ it
feels. The $Z_2$ symmetry imposes $r_{-s,h} = r_{s,-h}$; therefore all
flipping probabilities can be encoded in the flipping probability for
a $s=+1$ spin, $r_h \equiv r_{+1,h}$. The absence of bulk noise
imposes $r_{4}=0$.  The standard linear VM is given by $r_h =1/2 -
h/8$.  In the general case, the NLV model depends on four free
parameters, $r_{-4}$, $r_{-2}$, $r_0$ and $r_2$. In the following, we
adopt the arbitrary parametrization of
Ref.~\cite{PhysRevLett.87.045701}, imposing $r_{-2}=0.275$, $r_0=1/2$,
$r_{2} = r_{-4}/4$, and taking $r_{-4}\equiv \eps$ as a free tuning
parameter.
With this parametrization, the GV point corresponds to a critical
value $\eps_c$ separating a paramagnetic phase for $\eps > \eps_c$
from a ferromagnetic phase at $\eps < \eps_c$. 

\subsection{Critical point and critical exponents}
\label{sec:crit-point-crit}

As a first step in our analysis of the NLV model, we first check
  the results of Ref.~\cite{PhysRevLett.87.045701} by numerically
  evaluating its critical GV point, and estimating the value of the
  corresponding critical exponents.  The critical point can be
  estimated by monitoring the density of interfaces $\rho(t)$ and
  identifying $\eps_c$ as the value leading to a logarithmic decay
  $\rho(t) \sim 1/\ln(t)$, separating a constant behavior ($\eps >
  \eps_c$) from an algebraic decay ($\eps <
  \eps_c$)~\cite{PhysRevLett.87.045701}. Here we propose a different
  approach to 
  determine with high precision the critical point $\eps_c$, based on
  the behavior of the exit probability $E(x)$.  Indeed, $\eps>\eps_c$
  corresponds to a disordered paramagnetic phase with, for
  asymptotically large systems, $E(x) = 1/2$, while $\eps<\eps_c$
  corresponds to an ordered ferromagnetic phase, where $E(x)
  =\Theta(x)$, the Heaviside theta function.  Therefore, focusing on
  an initial density $x<1/2$, we should observe $E(x) \to 1/2$ for
  $\eps>\eps_c$, $E(x) \to 0$ for $\eps<\eps_c$, and $E(x) \to
  \mathrm{const} < 1/2$ for $\eps=\eps_c$ when increasing the system
  size $L$.

\begin{figure}[t]
  \centerline{\includegraphics[width=8cm]{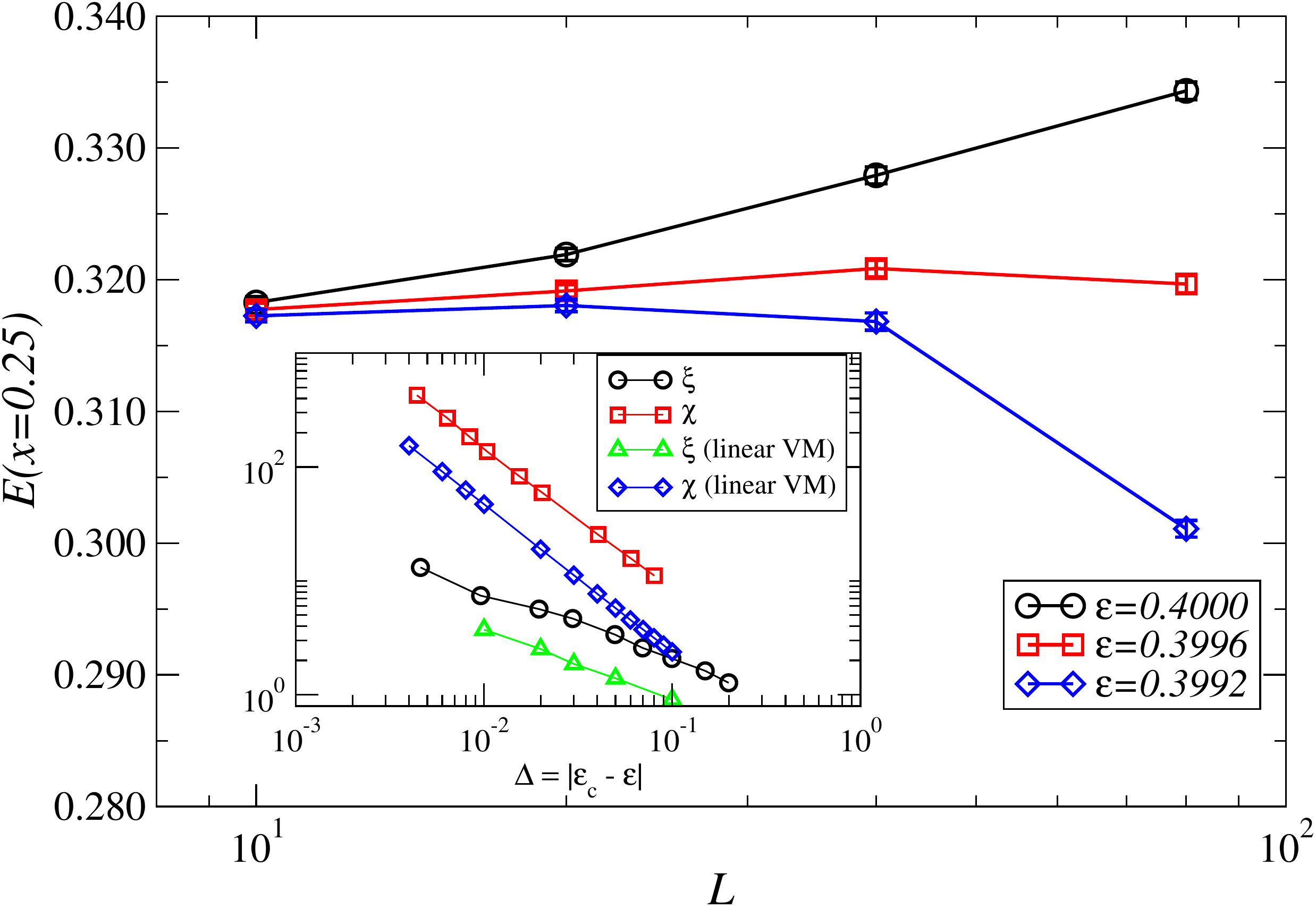}}
  \caption{(Color online) Main: Finite-size scaling of the exit
    probability at $x=0.25$ as a function of system size in the NLV
    model in $d=2$, computed over at least $5 \times 10^5$ independent
    realizations of the dynamics for each system size.  Inset:
    Susceptibility and correlation length as a function of $\Delta =
    |\eps_c - \eps|$ on lattices of size $L=100$ and $L=5000$,
    respectively.}
\label{fig:exitNLV}
\end{figure}

In Fig.~\ref{fig:exitNLV} (main plot) we report the exit probability
for $x=0.25$ and different values of $\eps$ as a function of the
lattice size $L$. A plateau is obtained for $\eps \simeq 0.3996$,
while larger (smaller) values of $\eps$ lead to an increase (decrease)
of $E(x=0.25)$ with $L$. We conclude that the critical point of the GV
transition in $d=2$ for the NLV model is located at $\eps_c =
0.3996(4)$, in good agreement with the result inferred from Fig.~3(a)
in Ref.~\cite{PhysRevLett.87.045701}, where a behavior for the density
of interfaces with the logarithmic VM form was found.

The properties of the GV universality class can be further explored by
considering several critical exponents, measured in the vicinity of
the critical point $\eps_c$. These exponents are usually defined in
terms of the susceptibility, measured as the fluctuations of the
magnetization $\phi=\sum_i s_i/N$, i.e.
\begin{equation}
  \label{eq:3}
  \chi = L^2\left[\av{\phi^2} - \av{|\phi|}^2\right],
\end{equation}
when approaching the transition from the paramagnetic, disordered
phase, and the correlation length $\xi$ which, when approaching the GV
manifold from the ferromagnetic, ordered phase, can be measured from
the relation \cite{PhysRevLett.87.045701}
\begin{equation}
  \label{eq:1}
  \rho(t) \sim \xi t^{-1/2}.
\end{equation}
Close to the critical point, these two quantities depend on $\Delta
= |\eps_c - \eps|$, defining the critical exponents
\begin{equation}
  \label{eq:4}
  \chi(\Delta) \sim \Delta^{-\gamma}, \qquad \xi(\Delta) \sim
  \Delta^{-\nu}.
\end{equation}

In Fig.~\ref{fig:exitNLV} (inset) we present the results of numerical
simulations of the quantities $\xi$ and $\chi$ as a function of
$\Delta$. While the determination of $\gamma$ is straightforward, the
measurement of $\nu$ is hindered by extremely long pre-asymptotic
effects in the curvature-driven regime, leading to a decay of
$\rho(t)$ with an effective numerical exponent smaller than
$1/2$~\cite{PhysRevE.78.011109}.  Here, in order to obtain information
about $\xi$, we proceed by performing a linear regression of
$1/\rho^2(t)$ as a function of $t$, and assigning to $\xi$ the value
of the slope thus obtained.  Data obtained in this way
(Fig.~\ref{fig:exitNLV} (inset)) provide the exponent values
$\nu\simeq 0.60$, $\gamma \simeq 1.26$.  The value of $\gamma$ is in
excellent agreement with early numerical
values~\cite{Deoliveira93,PhysRevLett.87.045701}, while $\nu$ is
rather different from the estimate of Dornic et
al.~\cite{PhysRevLett.87.045701} and compatible with the result from
\cite{Deoliveira93}.  Both exponents are also quite compatible with
the scaling relation $\gamma=2\nu$. With respect to the mean-field
values ($\nu=1/2$, $\gamma=1$, with logarithmic corrections) proposed
in Ref.~\cite{PhysRevLett.87.045701}, from our data it is difficult to
make a definite discrimination for the exponent $\nu$, since the plot
of $\xi$ as a function of $\Delta$ can be equally well fitted to a
pure power-law with non mean-field exponent or to a mean-field value
with logarithmic corrections. On the other hand, the exponent
$\gamma$ seems apparently better fitted with a non mean-field power-law
exponent.

To check these results we consider the linear VM in the GV manifold,
which in the NLV model introduced in \cite{PhysRevLett.87.045701} can
be approached by setting $r_{-4}=1$, $r_0=1/2$ and taking $r_{-2}
\equiv \eps \to \eps_c = 3/4$.  Here the state is paramagnetic for
$\eps< \eps_c$, while it is ferromagnetic for $\eps>\eps_c$.  In this
case, see Fig.~\ref{fig:exitNLV} (inset), we obtain
$\gamma_\mathrm{VM} \simeq 1.29$ and $\nu_\mathrm{VM} \simeq 0.62$,
which confirm the universality of the GV class. 

\subsection{Exit probability and consensus time}
\label{sec:exit-prob-cons}

\begin{figure}[t]
  \centering \includegraphics[width=8cm]{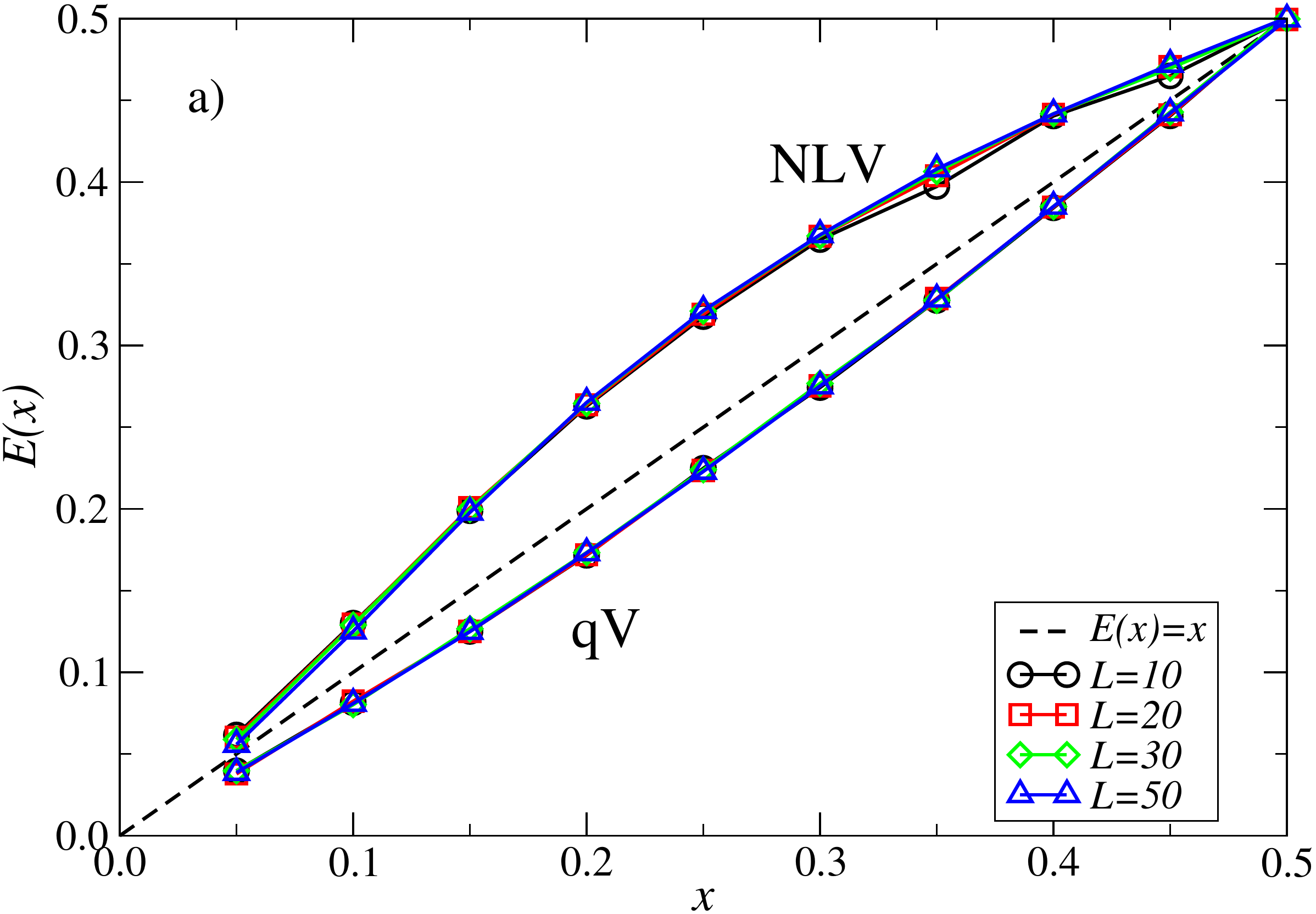}
  \includegraphics[width=8cm]{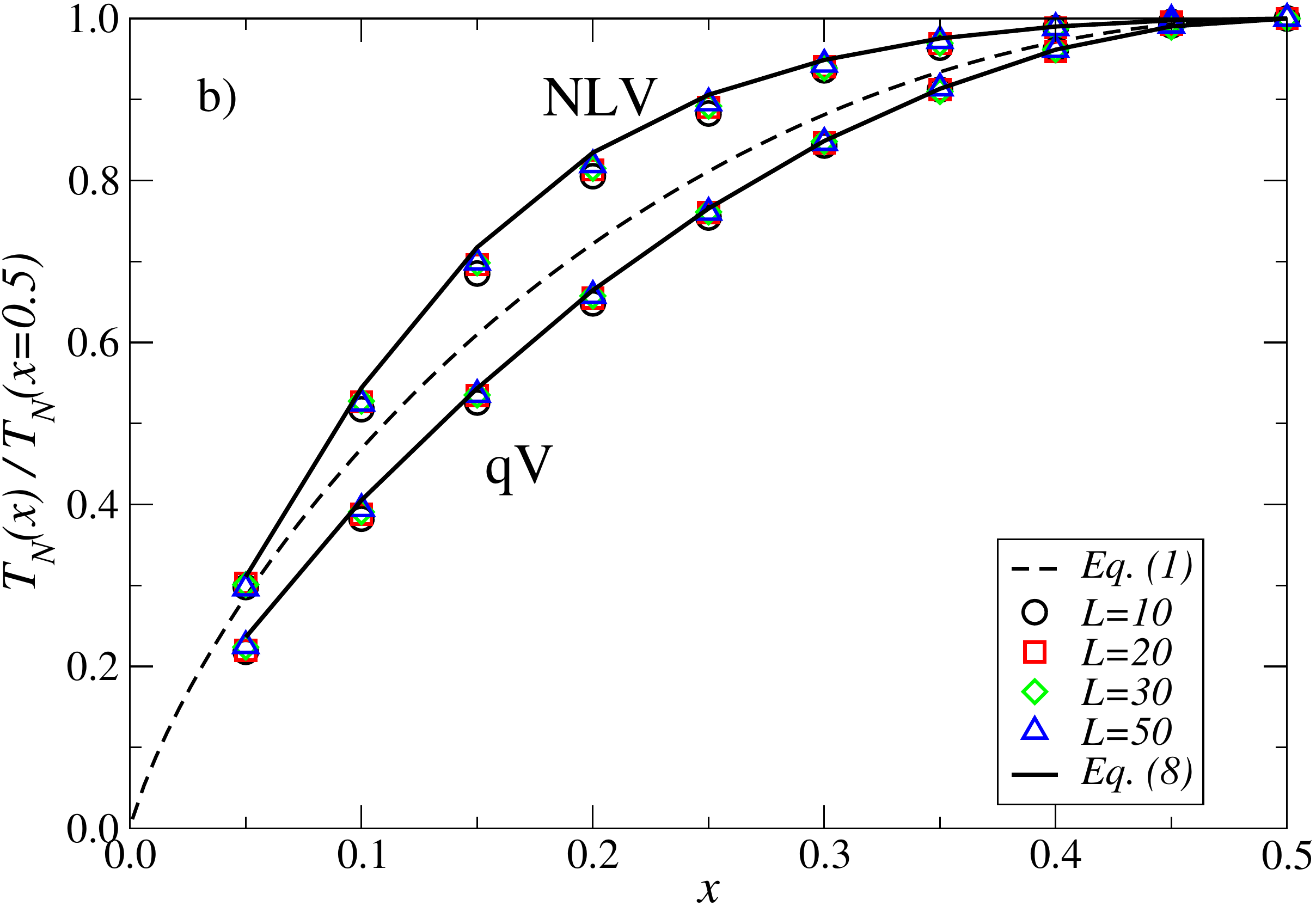}
  \caption{(Color online) (a) Exit probability as a function of the
    initial density for different system sizes in the NLV model and the
    $q=4$ qV model at the respective GV critical
    points. (b) Normalized consensus time as a function of $x$ for the
    same models.}
  \label{fig:exittimeNLV}
\end{figure}

The analysis presented above confirms the results of previous
studies. However, it also points out a new and surprising feature, which is
only evident in simulations performed out of the initial symmetric state
($x=1/2$).
As we can see from Fig.~\ref{fig:exitNLV} (main plot),
the exit probability of the NLV model at criticality (on the GV
manifold) computed at the non-symmetric homogeneous initial state
$x=0.25$, takes a value $E(x=0.25) = 0.32 \pm 0.02$, i.e. larger than
$0.25$, well beyond error bars. This observation hints towards a
non-linear form of the exit probability, which is confirmed in
Fig.~\ref{fig:exittimeNLV}a).
In this plot we can see that the exit probability $E(x)$ deviates from
linearity for the whole range of values of $x$. This deviation from
linear VM behavior extends also to the consensus time as a function of
$x$, as we can also see in Fig.~\ref{fig:exittimeNLV}b). 

\begin{figure}[t]
  \centerline{\includegraphics[width=8cm]{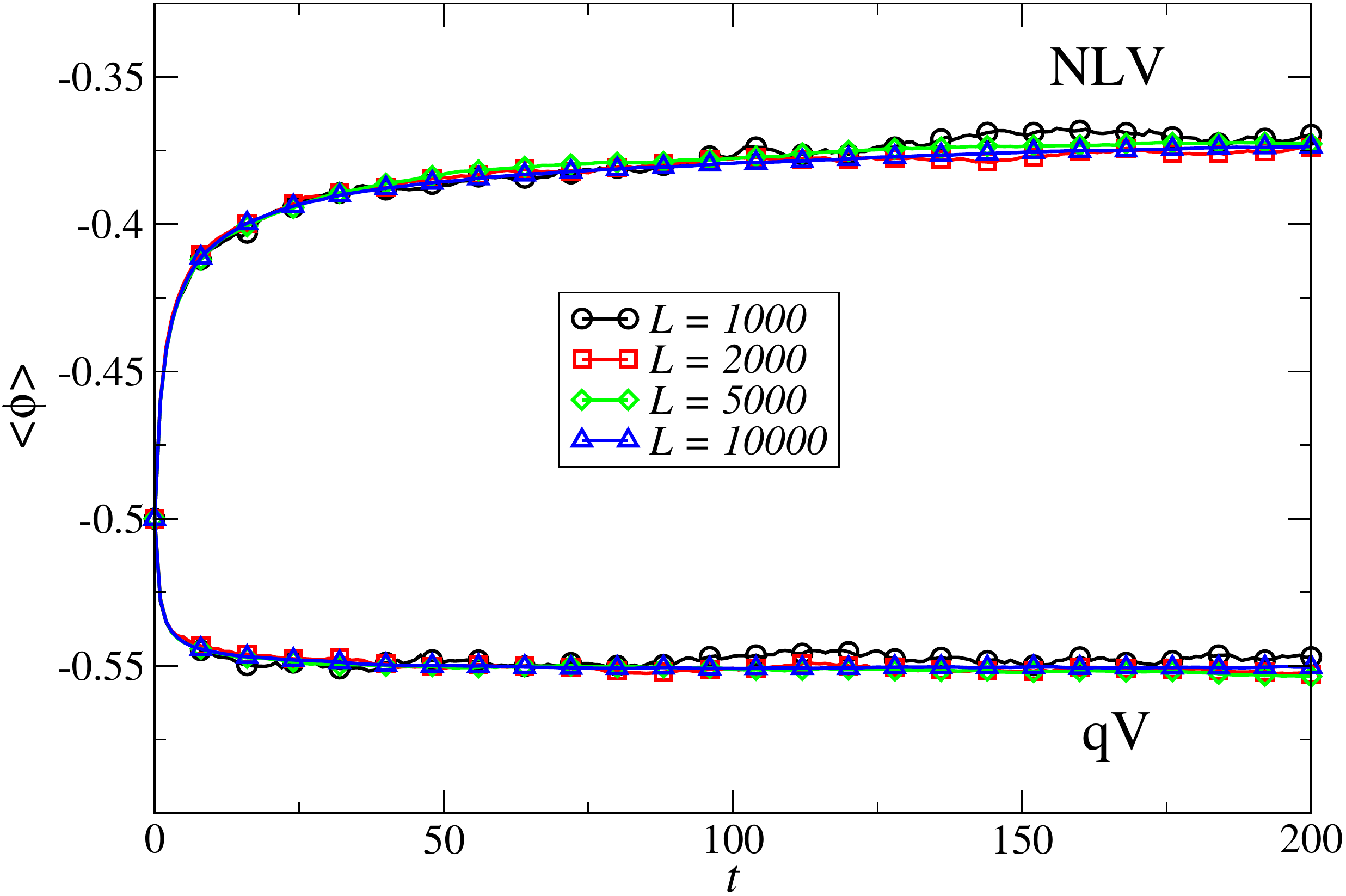}}
  \caption{(Color online) Average magnetization $\phi(t)$ as a
    function of time for different system sizes in the NLV model and
    the $q=4$ qV model at the respective GV critical points, for
    initial conditions with magnetization $\phi(0)=-0.5$,
    corresponding to $x=0.25$.}
  \label{fig:magNLV}
\end{figure}

\begin{figure}[t]
  \centerline{\includegraphics[width=8cm]{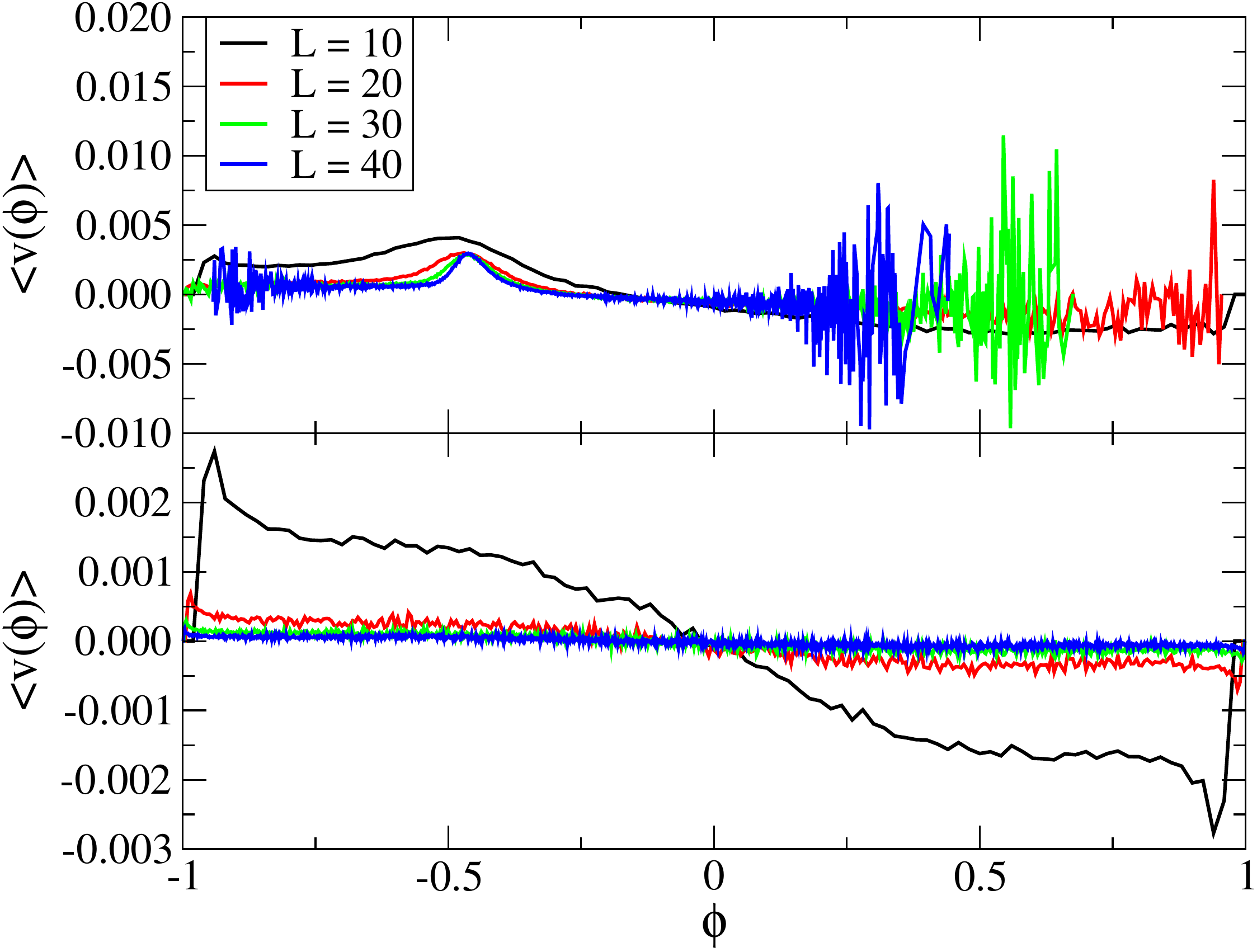}} 

  \caption{(Color online) Average drift as a function of the
    magnetization for different system sizes in the NLV model in $d=2$
    at the critical point, for system starting at an initial
    magnetization $\phi(0)=-0.5$ ($x=0.25$).  Top: $t<100$. Bottom:
    $t>100$.}
  \label{fig:driftNLV}
\end{figure}

This departure of the NLV model from linear VM behavior can be
understood by looking in detail at the time evolution of the
magnetization $\phi(t)$ in the system, starting from $x(0)=0.25$,
corresponding to $\phi(0)=2x(0)-1=-0.5$, see Fig.~\ref{fig:magNLV}.
Data shows that magnetization \textit{is strongly not conserved} at
short times, but in fact it experiences a sharp increase until it
stabilizes, for times $t \geq 100$, at a plateau with approximate
value $\phi_\infty \approx -0.37$. The peculiar time evolution of the
average magnetization (already noticed in
Ref.~\cite{PhysRevLett.87.045701}) can be related to the drift
$v(\phi)$ in a Langevin representation \cite{AlHammal05}, in the form
$\partial_t \av{\phi} = \av{v(\phi)}$ \cite{gardiner4ed2010}.  We
estimate the average drift $\av{v(\phi)}$ by computing $-2/N \sum_{i}
s_{i} f(x_{i})$ (where $x_{i}$ is the fraction of neighbors of $i$ in
opposite state) and we average all values of drift with the same
magnetization value $\phi=1/N \sum_{i} s_{i}$.  In
Fig.~\ref{fig:driftNLV} we plot $\av{v(\phi)}$ vs $\phi$ for different
system sizes in two distinct temporal regimes.  For short times
($t<100$) a sharp rise is present in the vicinity of the initial
magnetization. This is responsible for the initial increase of
magnetization until it reaches the steady state and its conserved (in
average) value.  For larger values of time $t>100$ this rise is
absent, and the drift takes a flat, almost vanishing form, thus
ensuring conservation of magnetization.

Given the flipping probability $f(x)$, the origin of the rise for
short times can be understood by considering the initial uncorrelated
condition. In that case the average drift is given by
\begin{equation}
\langle v(\phi)\rangle_{t=0} =
- \frac{2}{N} \sum_s \sum_h ~~s ~~f\left(\frac{4-s h}{8}\right)
~~{\rm Prob}(s,h),
\label{eq:6}
\end{equation}
where $h$ is the local
field (assuming even values between $-4$ and $4$), 
${\rm Prob}(-1,h)=(1-x) \binom{4}{k} x^k (1-x)^{4-k}$, 
${\rm Prob}(+1,h)=x \binom{4}{k} x^k (1-x)^{4-k}$,
and $k=(4+h)/2$. In the case of the NLV model defined in
Ref.~\cite{PhysRevLett.87.045701}, the flipping probability of a $+1$
spin takes the form $f_+(1)=r_{-4}$, $f_+(3/4)=r_{-2}$,
$f_+(1/2)=r_0$, $f_+(1/4)=r_2$, and $f_+(0)=0$.  
Performing the summations in Eq.~\eqref{eq:6} we obtain the drift as a
function of magnetization
\begin{equation}
  \label{eq:8}
   N \langle v(\phi)\rangle_{t=0} =\frac{1}{8} \phi (1-\phi^2) 
   F(\phi),
\end{equation}
where 
\begin{equation}
  \label{eq:9}
  F(\phi)=2(1-\phi^2)(2r_{-2} - 3r_0) + (3+\phi^2)(r_{-4} -
  4r_2).
\end{equation}
Depending on whether $F(\phi)$ is positive or negative, the drift will
have the same sign of $\phi$ or the opposite. For the NLV model, we
have $r_2 = r_{-4}/4$; therefore, in this case $F(\phi) =
2(1-\phi^2)(2r_{-2} - 3r_0)$ and it is negative for $r_{-2} < 3 r_0
/2$. For the values chosen for the NLV model, this inequality is indeed
satisfied, and therefore the initial drift is positive for $\phi<0$
($x<1/2$), negative for $\phi>0$ ($x>1/2$) and vanishes for $\phi=0$
($x=1/2$).

These results clarify the origin of the nonlinear exit probability
$E(x)$ in $Z_2$-symmetric models.  Initial conditions $x \neq 1/2$
imply a strong nonzero drift, which rapidly brings the fraction of
initial spins from its initial value $x$ to a different value $x'$.
After this short transient the build up of spatial correlations
cancels the drift, magnetization is conserved and the dynamics becomes
identical to that of linear VM.  As a consequence
$E_{NLV}(x)=E_{VM}(x')=x'$ as witnessed by Fig.~\ref{fig:magNLV},
where the density of $+1$ spins, starting from initial conditions
$x=0.25$, reaches a plateau $x' =(1+\phi_\infty)/2 \simeq 0.315$, in
good agreement with the estimate of the exit probability,
i.e. $E(0.25) \simeq 0.32$.  A similar mechanism is at the origin of
nonlinear exit probabilities for opinion dynamics models in
$d=1$~\cite{Castellano11}.  The same argument allows also to estimate
the deviation of the consensus time from the entropic form
Eq.~\eqref{eq:2}. In the short initial transient the fraction of $+1$
spins quickly converges to $x'=E(x)$.  The consensus time is
essentially set by the subsequent slow ordering, which occurs as in
the linear VM, hence we can write
\cite{PhysRevLett.94.178701,PhysRevE.82.010103,1751-8121-43-38-385003}
\begin{equation}
  T_{NLV}(x) = T_{VM}[E(x)].
  \label{eq:5}
\end{equation}
Figure~\ref{fig:exittimeNLV}b) confirms the correctness of this
theoretical estimate, and hints towards the relevance of the
microscopic details of the model, which induce a strong
drift at short time scales and lead in this way to nonuniversal
features such as a nonlinear exit probability and anomalous consensus time.

\section{The $q$-voter model}
\label{sec:q-voter-model}

In order to confirm the departure of the exit probability and of the
consensus time from linear voter behavior for $Z_2$-symmetric models
in the GV class, we consider the recently introduced non-linear
$q$-voter (qV) spin model, which is defined as follows
\cite{PhysRevE.80.041129}: At each time step $t$, a random site $i$ is
chosen; additionally $q$ nearest neighbor sites of $i$ are also
randomly selected (allowing for repetition to simplify the analysis),
and their spins examined.  If all the $q$ neighbors are in the same
state, the spin at site $i$ takes their common value. Otherwise, the
spin at $i$ flips its state with probability $\eps$. In any case, time
is updated $t \to t + 1/N$.  With this definition the flipping
probability in the qV model takes the form $f(x_i,q,\eps) = x_i^q +
\eps \left[1-x_i^q - \left(1-x_i\right)^q\right]$, where we remind
that $x_i$ is the fraction of neighbors of site $i$ in the opposite
state.  Following simple mean-field arguments
\cite{PhysRevE.80.041129,AlHammal05,Vazquez08}, one can show the
existence of a critical point $\eps=\eps_c(q)$, corresponding to GV
behavior, separating a paramagnetic (disordered) phase for
$\eps>\eps_c$ from a ferromagnetic (ordered) phase at $\eps<\eps_c$.
In a $d=1$ lattice, the qV model can be exactly mapped to the model of
nonconservative voters proposed in Ref.~\cite{Lambiotte08}.  From
here, one observe voter behavior at $\eps_c=1/2$ for any value of $q$,
while values of $\eps \neq 1/2$ lead to ordering dynamics with a
non-trivial, non-linear exit probability. In the more interesting case
$d=2$, numerical evidence presented in Ref.~\cite{PhysRevE.80.041129}
for the case $q=4$ indicated the presence of a critical point at
$\eps_c \simeq 1/4$. For this value of $\eps$, evidence of voter
behavior was found in terms of the decay of the density of interfaces
and the scaling of the correlation function, both of which are fully
compatible with the VM results.

\subsection{Critical point and critical exponents}
\label{sec:crit-point-crit-1}

Following the lines of the analysis carried out for the NLV model
  in Sec.~\ref{sec:crit-point-crit}, we first determine precisely the
  critical point of $q=4$ qV model by performing a finite size scaling
  analysis of the exit probability at $x=0.25$.
\begin{figure}[t]
  \centerline{\includegraphics[width=7cm]{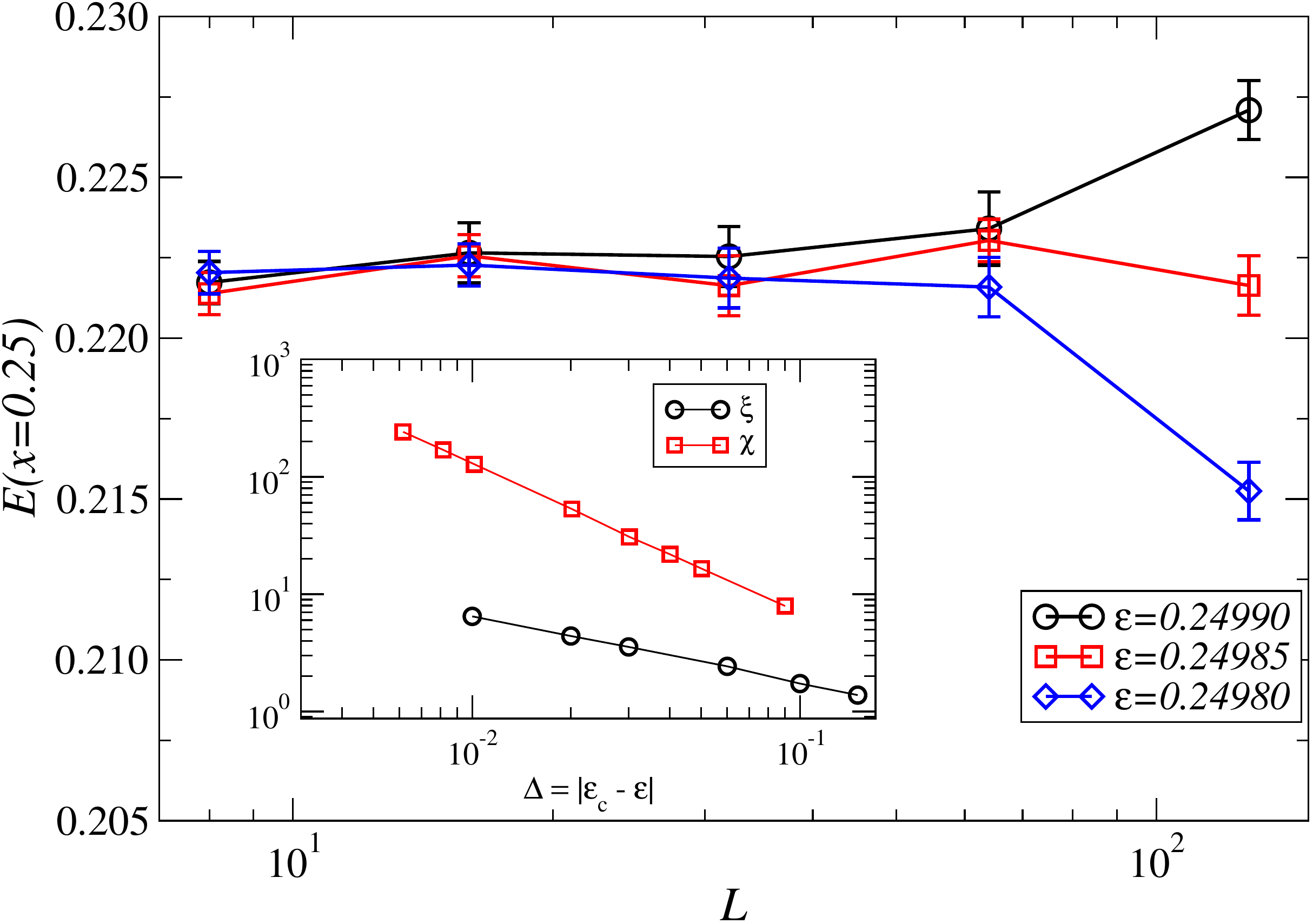}}
  \caption{(Color online) Main: Finite-size scaling of the exit
    probability at $x=0.25$ as a function of system size in the $q=4$
    qV model in $d=2$, computed over at least $5 \times 10^5$
    independent realizations of the dynamics for each system size.
    Inset: Susceptibility and correlation length as a function of
    $\Delta = |\eps_c - \eps|$ for the $q=4$ qV model on lattices of
    size $L=100$ and $L=5000$, respectively.}
\label{fig:exitq=4}
\end{figure}
In Fig.~\ref{fig:exitq=4} (main plot) we report the exit probability
for this value of $x$ and different values of $\eps$ as a function of
the lattice size $L$. A plateau is obtained for $\eps \simeq 0.24985$,
while larger (smaller) values of $\eps$ lead to an increase (decrease)
of $E(x=0.25)$ with $L$. We conclude that the critical point of the GV
transition in $d=2$ for the $q=4$ qV model is located at $\eps_c =
0.24985(5)$, in good agreement with the previous estimate in
Ref.~\cite{PhysRevE.80.041129}.  We further confirm the fact that the
critical $q=4$ qV model belongs to the GV class by computing the
exponents $\nu$ and $\gamma$, Fig.~\ref{fig:exitq=4} (inset). We
obtain the values $\nu \simeq 0.57$ and $\gamma \simeq 1.28$, in
reasonable agreement with our estimates for the NLV model.

\subsection{Exit probability and consensus time}
\label{sec:exit-prob-cons-1}

As in the case of the NLV model, the exit probability at $x=0.25$
indicates the presence of a nonlinear form, taking a value $E(x=0.25)=
0.222\pm0.007$, smaller than $0.25$ beyond the estimated error
bars. The non-linearity of $E(x)$ is further confirmed in
Fig.~\ref{fig:exittimeNLV}a), where we compare the function $E(x)$ for
$x<0.5$ with the linear form valid for linear VM. $E(x)$ is non-linear
in the whole range of $x$ values, being independent of $L$ at this
critical point $\eps_c$.  The deviation of the qV model from linear VM
behavior extends, similarly to the NLV model, to the functional
dependence with $x$ of the consensus time $T_N(x)$ at the critical
point $\eps_c$, as shown in Fig.~\ref{fig:exittimeNLV}b).

Noticeably, in the case of the qV model, the exit probability is
smaller than $x$, in opposition to the NLV model, where we observed
values $E(x)>x$. This smaller value of the exit probability is
reflected in the evolution of the magnetization $\phi(t)$, see
Fig.~\ref{fig:magNLV}, which is again strongly not conserved at short
times, exhibiting a sharp drop until it stabilizes, for times $t \geq
50$, at a plateau with approximate value $\phi_\infty \approx -0.55$.
\begin{figure}[t]
  \centerline{\includegraphics[width=8cm]{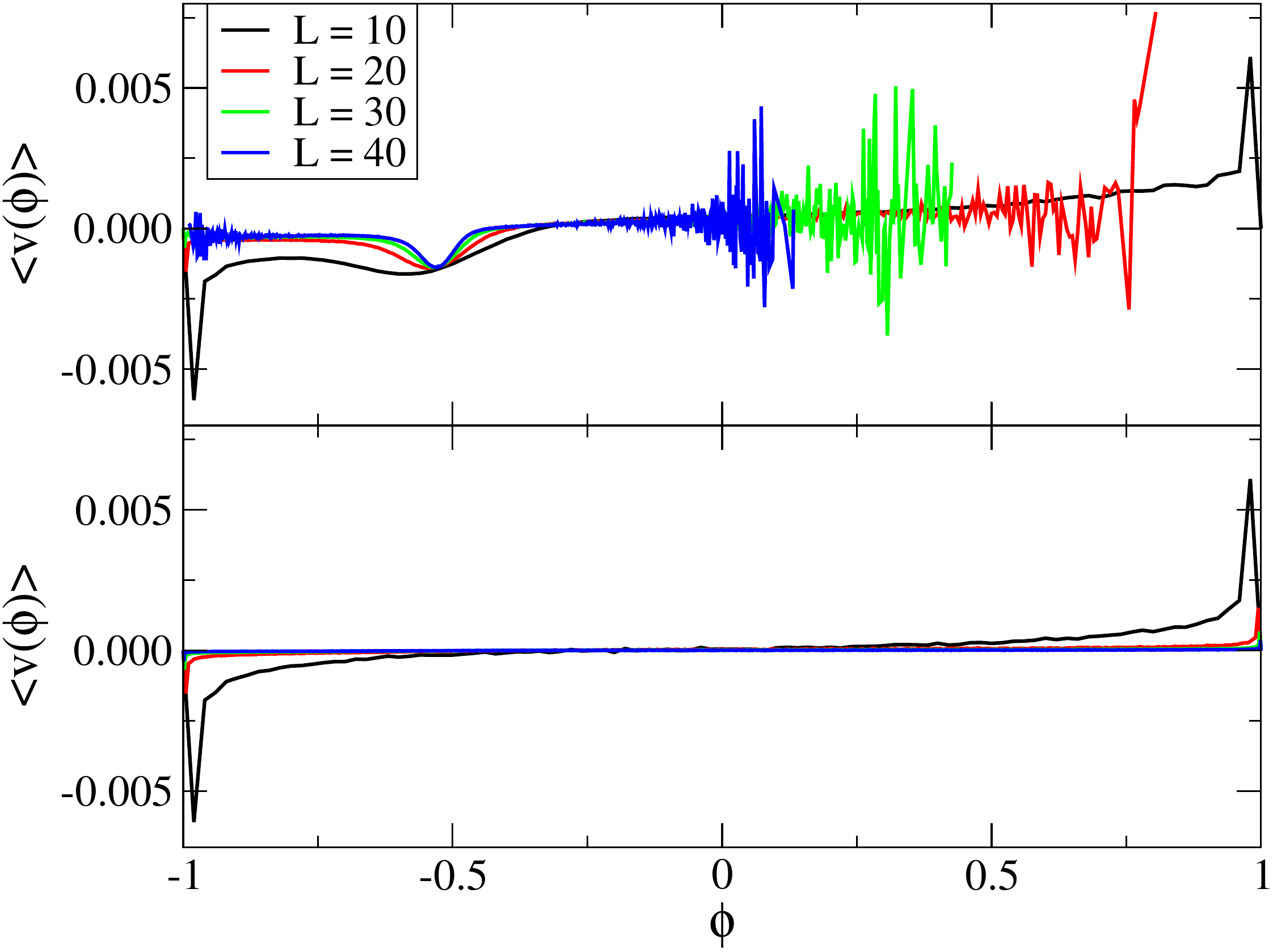}} 

  \caption{ (Color online) Average drift as a function of the
    magnetization for different system sizes in the $q=4$ voter model
    in $d=2$ at the critical point, for system starting at an initial
    magnetization $\phi(0)=-0.5$ ($x=0.25$).  Top: $t<50$. Bottom:
    $t>50$. }
  \label{fig:driftq=4}
\end{figure}
The time evolution of the average magnetization is again related with the
drift $v(\phi)$. 
In Fig.~\ref{fig:driftq=4} we plot $\av{v(\phi)}$ vs $\phi$ for different
system sizes in two distinct temporal regimes.
For short times ($t<50$) a sharp dip is present in the vicinity of
the initial magnetization. This is responsible for the initial
decrease of magnetization until it reaches the steady state and its
conserved (in average) value. 
For larger values of time ($t>50$) this dip is absent,
and the drift takes a flat, almost vanishing form, thus ensuring
conservation of magnetization.

The dip in the drift a short times can also be understood by computing
the drift in the initial uncorrelated condition. From
Eqs.~\eqref{eq:8} and~\eqref{eq:9}, and considering that for the qV
model the flipping rates $r_i$ can be written as $r_{-4}=1$, $r_{-2} =
3 (27 + 58 \eps)/256$, $r_0 = (1 + 14 \eps)/16$, $r_2 = (1 + 174
\eps)/256$, and $r_4=1$, we obtain
\begin{equation}
  \label{eq:10}
  F(\phi) = \frac{3}{32} \left[(1-2\eps)\phi ^2 + 41-114\eps \right],
\end{equation}
which is a positive function for all $\phi$ for $\eps<41/114$. Thus,
for $\eps=\eps_c$, we find that the initial drift is negative for
$\phi<0$ ($x<1/2$), positive for $\phi>0$ ($x>1/2$) and vanishes for
$\phi=0$ ($x=1/2$).  Again, for the qV model the nonlinear exit
probability and the anomalous consensus time can be related through
the argument leading to Eq.~\eqref{eq:5}. Indeed, for $x=0.25$, from
Fig.~\ref{fig:magNLV} we read a stationary large time magnetization
$\phi_\infty \approx -0.55$, corresponding to $x' \simeq 0.225$, is in
good agreement with the estimate of the exit probability,
i.e. $E(0.25) \simeq 0.222$. Eq.~\eqref{eq:5} is again valid for the
whole range of values of $x$, as shown in Fig.~\ref{fig:exittimeNLV}.

\section{Conserved magnetization: the Kaya, Kabak{\c c}io{\v g}lu, and
  Erzan model}
\label{sec:kaya-kabakc-ciov}

\begin{figure}[t]
  \centerline{\includegraphics[width=8cm]{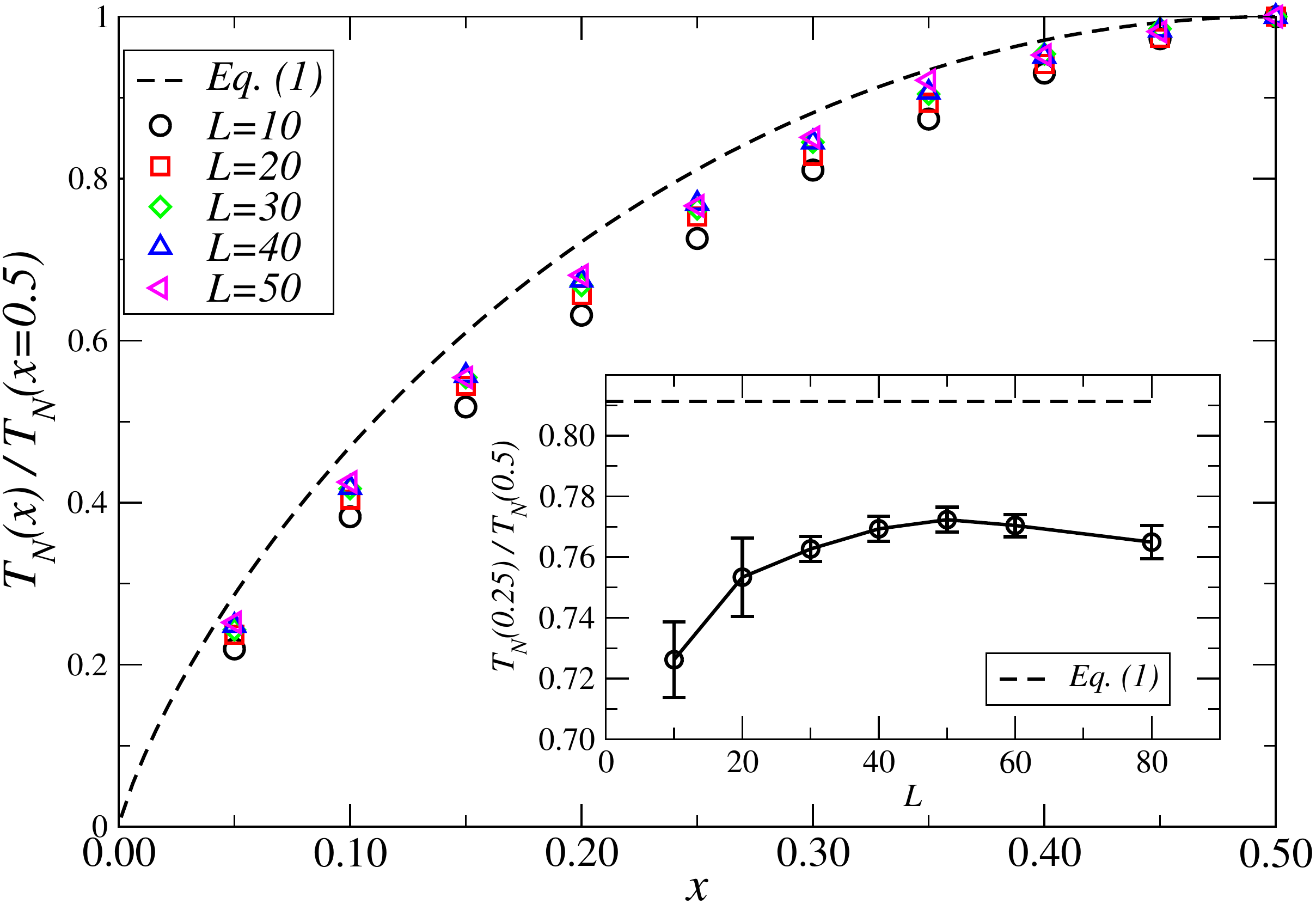}}
  \caption{(Color online) Main plot: Consensus time as a function of
    $x$ for different system sizes in the KKE model at the
    delocalization transition in $d=2$.  Inset: Evolution of
    $T_N(x)/T_N(0.5)$ for $x=0.25$ as a function of the system size
    $L$.}
\label{fig:timeexitKKE}
\end{figure}

From the analysis of the results obtained for the NLV and qV models, we
expect that any $Z_2$-symmetric model belonging to the GV class and
endowed with a nonlinear flipping probability $f(x_i)$ will exhibit a
strictly nonlinear exit probability and a ``non-entropic'' form of the
consensus time $T_N(x)$.  By the same token, it is clear that in other
type of models belonging to the GV class, conservation of the average
magnetization will guarantee a linear exit
probability~\cite{KineticViewRedner}, as for linear VM.  The question
naturally arises about the form of the consensus times in those models
in the GV class in which conservation of magnetization is enforced.

To answer this question we have considered the KKE interface model at
the delocalization transition \cite{KKEmodel00}, which can be
formulated in terms of a spin systems as follows: At each time step a
spin $s_{i,j}$ and one of its neighbors are randomly selected; if they
are equal nothing happens, otherwise the number $n^+$ of positive
neighbors of the negative spin is computed and, with probability
$1/n^+$, $s_{i,j}$ and the neighbor are made equal.  This model
belongs to the GV class, as has been shown in
Ref.~\cite{PhysRevLett.87.045701}, where a logarithmic decay of the
density of interfaces was observed.  In Fig.~\ref{fig:timeexitKKE} we
plot the rescaled consensus time as a function of $x$ for different
system sizes. These results indicate that again the consensus time
deviates from the form expected in linear VM. The deviation from the
behavior given by Eq.~\eqref{eq:2} should be attributed again to
relevant microscopic details of the model. However, the mechanism
inducing the deviation is here necessarily different from the one
responsible for the deviation for the NLV and qV models (i.e. initial
short timescale nonconservation of magnetization).  Its investigation
constitutes an interesting line for future research.

\section{Conclusions}
\label{sec:conclusions}

In summary, in this paper we have shown, using large scale numerical
simulations of three different spin models, that different subclasses
of the generalized voter class for ordering dynamics actually exhibit
different behaviors when unbalanced initial conditions are considered.
All elements of the GV class are broadly characterized by lack of
surface tension and a logarithmic decay of the density of interfaces
in $d=2$. However, when looking at the exit probability and the
consensus time, two types of behavior occur.  In one subclass,
encompassing systems, such as KKE, with no $Z_2$ symmetry but
magnetization conserved on average, the consensus time differs from
the entropic form characterizing the VM, while the exit probability is
linear.  In the other subclass, composed by systems with $Z_2$
symmetry and nonlinear form of the flipping probability $f(x)$, both
the exit probability and the consensus time differ from the VM.  This
variation is essentially due to a nonconservation of magnetization
during a short initial transient.  The buildup of spatial correlations
rapidly leads to an effective cancellation of the drift, so that the
subsequent evolution is the same as for the linear VM, but starting
from $x' \neq x$ so that $E(x)$ and $T(x)$ are modified.  To the best
of our knowledge the results for the NLV and qV models reported here
constitute the first example of a dynamics in $d=2$ with an exit
probability different (in the large system size limit) from the
step-function (typical of dynamics driven by surface tension) or the
linear shape of VM. Nontrivial shapes were previously found only in
$d=1$~\cite{Lambiotte08,Slanina08} and in mean-field, but not in
$d=2$.  Finally, our numerical estimates of the critical exponents
associated to the GV transition suggest a value of the $\gamma$
exponent possibly different from the mean-field value previously
supposed to hold.  Further research work is needed to fully clarify
this issue, based on numerical simulations on systems sizes
  beyond those used in the present work, which are at the boundary of
  our computational limits.

\begin{acknowledgments}
  R.P.-S. acknowledges financial support from the Spanish MICINN,
  under project FIS2010-21781-C02-01; ICREA Academia, funded by the
  Generalitat de Catalunya; and the Junta de Andaluc\'{i}a, under
  project No. P09-FQM4682.  C.C. acknowledges financial support from
  the European Science Foundation, under project DRUST.
\end{acknowledgments}


\begin{thebibliography}{29}%
\makeatletter
\providecommand \@ifxundefined [1]{%
 \@ifx{#1\undefined}
}%
\providecommand \@ifnum [1]{%
 \ifnum #1\expandafter \@firstoftwo
 \else \expandafter \@secondoftwo
 \fi
}%
\providecommand \@ifx [1]{%
 \ifx #1\expandafter \@firstoftwo
 \else \expandafter \@secondoftwo
 \fi
}%
\providecommand \natexlab [1]{#1}%
\providecommand \enquote  [1]{``#1''}%
\providecommand \bibnamefont  [1]{#1}%
\providecommand \bibfnamefont [1]{#1}%
\providecommand \citenamefont [1]{#1}%
\providecommand \href@noop [0]{\@secondoftwo}%
\providecommand \href [0]{\begingroup \@sanitize@url \@href}%
\providecommand \@href[1]{\@@startlink{#1}\@@href}%
\providecommand \@@href[1]{\endgroup#1\@@endlink}%
\providecommand \@sanitize@url [0]{\catcode `\\12\catcode `\$12\catcode
  `\&12\catcode `\#12\catcode `\^12\catcode `\_12\catcode `\%12\relax}%
\providecommand \@@startlink[1]{}%
\providecommand \@@endlink[0]{}%
\providecommand \url  [0]{\begingroup\@sanitize@url \@url }%
\providecommand \@url [1]{\endgroup\@href {#1}{\urlprefix }}%
\providecommand \urlprefix  [0]{URL }%
\providecommand \Eprint [0]{\href }%
\providecommand \doibase [0]{http://dx.doi.org/}%
\providecommand \selectlanguage [0]{\@gobble}%
\providecommand \bibinfo  [0]{\@secondoftwo}%
\providecommand \bibfield  [0]{\@secondoftwo}%
\providecommand \translation [1]{[#1]}%
\providecommand \BibitemOpen [0]{}%
\providecommand \bibitemStop [0]{}%
\providecommand \bibitemNoStop [0]{.\EOS\space}%
\providecommand \EOS [0]{\spacefactor3000\relax}%
\providecommand \BibitemShut  [1]{\csname bibitem#1\endcsname}%
\let\auto@bib@innerbib\@empty
\bibitem [{\citenamefont {Clifford}\ and\ \citenamefont
  {Sudbury}(1973)}]{Clifford73}%
  \BibitemOpen
  \bibfield  {author} {\bibinfo {author} {\bibfnamefont {P.}~\bibnamefont
  {Clifford}}\ and\ \bibinfo {author} {\bibfnamefont {A.}~\bibnamefont
  {Sudbury}},\ }\href@noop {} {\bibfield  {journal} {\bibinfo  {journal}
  {Biometrika}\ }\textbf {\bibinfo {volume} {60}},\ \bibinfo {pages} {581}
  (\bibinfo {year} {1973})}\BibitemShut {NoStop}%
\bibitem [{\citenamefont {Holley}\ and\ \citenamefont
  {Liggett}(1975)}]{Holley:1975fk}%
  \BibitemOpen
  \bibfield  {author} {\bibinfo {author} {\bibfnamefont {R.~A.}\ \bibnamefont
  {Holley}}\ and\ \bibinfo {author} {\bibfnamefont {T.~M.}\ \bibnamefont
  {Liggett}},\ }\href@noop {} {\bibfield  {journal} {\bibinfo  {journal}
  {Annals of Probability}\ }\textbf {\bibinfo {volume} {3}},\ \bibinfo {pages}
  {643} (\bibinfo {year} {1975})}\BibitemShut {NoStop}%
\bibitem [{\citenamefont {Bray}(1994)}]{Bray94}%
  \BibitemOpen
  \bibfield  {author} {\bibinfo {author} {\bibfnamefont {A.~J.}\ \bibnamefont
  {Bray}},\ }\href@noop {} {\bibfield  {journal} {\bibinfo  {journal} {Adv.
  Phys.}\ }\textbf {\bibinfo {volume} {43}},\ \bibinfo {pages} {357} (\bibinfo
  {year} {1994})}\BibitemShut {NoStop}%
\bibitem [{\citenamefont {Krapivsky}\ \emph {et~al.}(2010)\citenamefont
  {Krapivsky}, \citenamefont {Redner},\ and\ \citenamefont
  {{Ben-Naim}}}]{KineticViewRedner}%
  \BibitemOpen
  \bibfield  {author} {\bibinfo {author} {\bibfnamefont {P.}~\bibnamefont
  {Krapivsky}}, \bibinfo {author} {\bibfnamefont {S.}~\bibnamefont {Redner}}, \
  and\ \bibinfo {author} {\bibfnamefont {E.}~\bibnamefont {{Ben-Naim}}},\
  }\href@noop {} {\emph {\bibinfo {title} {A Kinetic View of Statistical
  Physics}}}\ (\bibinfo  {publisher} {Cambridge University Press},\ \bibinfo
  {address} {Cambridge},\ \bibinfo {year} {2010})\BibitemShut {NoStop}%
\bibitem [{\citenamefont {Liggett}(1999)}]{liggett99:_stoch_inter}%
  \BibitemOpen
  \bibfield  {author} {\bibinfo {author} {\bibfnamefont {T.~M.}\ \bibnamefont
  {Liggett}},\ }\href@noop {} {\emph {\bibinfo {title} {Stochastic interacting
  particle systems: Contact, Voter, and Exclusion processes}}}\ (\bibinfo
  {publisher} {Springer-Verlag},\ \bibinfo {address} {New York},\ \bibinfo
  {year} {1999})\BibitemShut {NoStop}%
\bibitem [{\citenamefont {Frachebourg}\ and\ \citenamefont
  {Krapivsky}(1996)}]{PhysRevE.53.R3009}%
  \BibitemOpen
  \bibfield  {author} {\bibinfo {author} {\bibfnamefont {L.}~\bibnamefont
  {Frachebourg}}\ and\ \bibinfo {author} {\bibfnamefont {P.~L.}\ \bibnamefont
  {Krapivsky}},\ }\href@noop {} {\bibfield  {journal} {\bibinfo  {journal}
  {Phys. Rev. E}\ }\textbf {\bibinfo {volume} {53}},\ \bibinfo {pages} {R3009}
  (\bibinfo {year} {1996})}\BibitemShut {NoStop}%
\bibitem [{\citenamefont {Castellano}\ \emph
  {et~al.}(2009{\natexlab{a}})\citenamefont {Castellano}, \citenamefont
  {Fortunato},\ and\ \citenamefont {Loreto}}]{Castellano09}%
  \BibitemOpen
  \bibfield  {author} {\bibinfo {author} {\bibfnamefont {C.}~\bibnamefont
  {Castellano}}, \bibinfo {author} {\bibfnamefont {S.}~\bibnamefont
  {Fortunato}}, \ and\ \bibinfo {author} {\bibfnamefont {V.}~\bibnamefont
  {Loreto}},\ }\href@noop {} {\bibfield  {journal} {\bibinfo  {journal} {Rev.
  Mod. Phys.}\ }\textbf {\bibinfo {volume} {81}},\ \bibinfo {pages} {591}
  (\bibinfo {year} {2009}{\natexlab{a}})}\BibitemShut {NoStop}%
\bibitem [{\citenamefont {Crow}\ and\ \citenamefont {Kimura}(1970)}]{Crowbook}%
  \BibitemOpen
  \bibfield  {author} {\bibinfo {author} {\bibfnamefont {J.~F.}\ \bibnamefont
  {Crow}}\ and\ \bibinfo {author} {\bibfnamefont {M.}~\bibnamefont {Kimura}},\
  }\href@noop {} {\emph {\bibinfo {title} {An introduction to population
  genetics theory}}}\ (\bibinfo  {publisher} {Harper \& Row},\ \bibinfo
  {address} {New York},\ \bibinfo {year} {1970})\BibitemShut {NoStop}%
\bibitem [{\citenamefont {Hubbell}(2001)}]{hubbell2001unified}%
  \BibitemOpen
  \bibfield  {author} {\bibinfo {author} {\bibfnamefont {S.}~\bibnamefont
  {Hubbell}},\ }\href@noop {} {\emph {\bibinfo {title} {The Unified Neutral
  Theory of Biodiversity and Biogeography}}},\ Monographs in Population
  Biology\ (\bibinfo  {publisher} {Princeton University Press},\ \bibinfo
  {address} {Princeton, NJ},\ \bibinfo {year} {2001})\BibitemShut {NoStop}%
\bibitem [{\citenamefont {Blythe}(2009)}]{Blythe09}%
  \BibitemOpen
  \bibfield  {author} {\bibinfo {author} {\bibfnamefont {R.~A.}\ \bibnamefont
  {Blythe}},\ }\href {http://stacks.iop.org/1742-5468/2009/i=02/a=P02059}
  {\bibfield  {journal} {\bibinfo  {journal} {Journal of Statistical Mechanics:
  Theory and Experiment}\ }\textbf {\bibinfo {volume} {2009}},\ \bibinfo
  {pages} {P02059} (\bibinfo {year} {2009})}\BibitemShut {NoStop}%
\bibitem [{\citenamefont {Dornic}\ \emph {et~al.}(2001)\citenamefont {Dornic},
  \citenamefont {Chat\'e}, \citenamefont {Chave},\ and\ \citenamefont
  {Hinrichsen}}]{PhysRevLett.87.045701}%
  \BibitemOpen
  \bibfield  {author} {\bibinfo {author} {\bibfnamefont {I.}~\bibnamefont
  {Dornic}}, \bibinfo {author} {\bibfnamefont {H.}~\bibnamefont {Chat\'e}},
  \bibinfo {author} {\bibfnamefont {J.}~\bibnamefont {Chave}}, \ and\ \bibinfo
  {author} {\bibfnamefont {H.}~\bibnamefont {Hinrichsen}},\ }\href@noop {}
  {\bibfield  {journal} {\bibinfo  {journal} {Phys. Rev. Lett.}\ }\textbf
  {\bibinfo {volume} {87}},\ \bibinfo {pages} {045701} (\bibinfo {year}
  {2001})}\BibitemShut {NoStop}%
\bibitem [{\citenamefont {Blythe}(2010)}]{1751-8121-43-38-385003}%
  \BibitemOpen
  \bibfield  {author} {\bibinfo {author} {\bibfnamefont {R.~A.}\ \bibnamefont
  {Blythe}},\ }\href@noop {} {\bibfield  {journal} {\bibinfo  {journal}
  {Journal of Physics A: Mathematical and Theoretical}\ }\textbf {\bibinfo
  {volume} {43}},\ \bibinfo {pages} {385003} (\bibinfo {year}
  {2010})}\BibitemShut {NoStop}%
\bibitem [{\citenamefont {de~Oliveira}\ \emph {et~al.}(1993)\citenamefont
  {de~Oliveira}, \citenamefont {Mendes},\ and\ \citenamefont
  {Santos}}]{Deoliveira93}%
  \BibitemOpen
  \bibfield  {author} {\bibinfo {author} {\bibfnamefont {M.}~\bibnamefont
  {de~Oliveira}}, \bibinfo {author} {\bibfnamefont {J.}~\bibnamefont {Mendes}},
  \ and\ \bibinfo {author} {\bibfnamefont {M.}~\bibnamefont {Santos}},\
  }\href@noop {} {\bibfield  {journal} {\bibinfo  {journal} {J. Phys. A}\
  }\textbf {\bibinfo {volume} {26}},\ \bibinfo {pages} {2317} (\bibinfo {year}
  {1993})}\BibitemShut {NoStop}%
\bibitem [{\citenamefont {{Drouffe}}\ and\ \citenamefont
  {{Godr{\`e}che}}(1999)}]{Drouffe99}%
  \BibitemOpen
  \bibfield  {author} {\bibinfo {author} {\bibfnamefont {J.-M.}\ \bibnamefont
  {{Drouffe}}}\ and\ \bibinfo {author} {\bibfnamefont {C.}~\bibnamefont
  {{Godr{\`e}che}}},\ }\href@noop {} {\bibfield  {journal} {\bibinfo  {journal}
  {J. Phys. A}\ }\textbf {\bibinfo {volume} {32}},\ \bibinfo {pages} {249}
  (\bibinfo {year} {1999})}\BibitemShut {NoStop}%
\bibitem [{\citenamefont {Molofsky}\ \emph {et~al.}(1999)\citenamefont
  {Molofsky}, \citenamefont {Durrett}, \citenamefont {Dushoff}, \citenamefont
  {Griffeath},\ and\ \citenamefont {Levin}}]{Molofsky99}%
  \BibitemOpen
  \bibfield  {author} {\bibinfo {author} {\bibfnamefont {J.}~\bibnamefont
  {Molofsky}}, \bibinfo {author} {\bibfnamefont {R.}~\bibnamefont {Durrett}},
  \bibinfo {author} {\bibfnamefont {J.}~\bibnamefont {Dushoff}}, \bibinfo
  {author} {\bibfnamefont {D.}~\bibnamefont {Griffeath}}, \ and\ \bibinfo
  {author} {\bibfnamefont {S.}~\bibnamefont {Levin}},\ }\href@noop {}
  {\bibfield  {journal} {\bibinfo  {journal} {Theoretical Population Biology}\
  }\textbf {\bibinfo {volume} {55}},\ \bibinfo {pages} {270 } (\bibinfo {year}
  {1999})}\BibitemShut {NoStop}%
\bibitem [{\citenamefont {{Al Hammal}}\ \emph {et~al.}(2005)\citenamefont {{Al
  Hammal}}, \citenamefont {Chat\'{e}}, \citenamefont {Dornic},\ and\
  \citenamefont {Mu{\~{n}}oz}}]{AlHammal05}%
  \BibitemOpen
  \bibfield  {author} {\bibinfo {author} {\bibfnamefont {O.}~\bibnamefont {{Al
  Hammal}}}, \bibinfo {author} {\bibfnamefont {H.}~\bibnamefont {Chat\'{e}}},
  \bibinfo {author} {\bibfnamefont {I.}~\bibnamefont {Dornic}}, \ and\ \bibinfo
  {author} {\bibfnamefont {M.~A.}\ \bibnamefont {Mu{\~{n}}oz}},\ }\href@noop {}
  {\bibfield  {journal} {\bibinfo  {journal} {Phys. Rev. Lett.}\ }\textbf
  {\bibinfo {volume} {94}},\ \bibinfo {pages} {230601} (\bibinfo {year}
  {2005})}\BibitemShut {NoStop}%
\bibitem [{\citenamefont {V\'{a}zquez}\ and\ \citenamefont
  {L\'{o}pez}(2008)}]{Vazquez08}%
  \BibitemOpen
  \bibfield  {author} {\bibinfo {author} {\bibfnamefont {F.}~\bibnamefont
  {V\'{a}zquez}}\ and\ \bibinfo {author} {\bibfnamefont {C.}~\bibnamefont
  {L\'{o}pez}},\ }\href@noop {} {\bibfield  {journal} {\bibinfo  {journal}
  {Phys. Rev. E}\ }\textbf {\bibinfo {volume} {78}},\ \bibinfo {pages} {061127}
  (\bibinfo {year} {2008})}\BibitemShut {NoStop}%
\bibitem [{\citenamefont {Canet}\ \emph {et~al.}(2005)\citenamefont {Canet},
  \citenamefont {Chat\'e}, \citenamefont {Delamotte}, \citenamefont {Dornic},\
  and\ \citenamefont {Mu\~noz}}]{PhysRevLett.95.100601}%
  \BibitemOpen
  \bibfield  {author} {\bibinfo {author} {\bibfnamefont {L.}~\bibnamefont
  {Canet}}, \bibinfo {author} {\bibfnamefont {H.}~\bibnamefont {Chat\'e}},
  \bibinfo {author} {\bibfnamefont {B.}~\bibnamefont {Delamotte}}, \bibinfo
  {author} {\bibfnamefont {I.}~\bibnamefont {Dornic}}, \ and\ \bibinfo {author}
  {\bibfnamefont {M.~A.}\ \bibnamefont {Mu\~noz}},\ }\href@noop {} {\bibfield
  {journal} {\bibinfo  {journal} {Phys. Rev. Lett.}\ }\textbf {\bibinfo
  {volume} {95}},\ \bibinfo {pages} {100601} (\bibinfo {year}
  {2005})}\BibitemShut {NoStop}%
\bibitem [{\citenamefont {Droz}\ \emph {et~al.}(2003)\citenamefont {Droz},
  \citenamefont {Ferreira},\ and\ \citenamefont {Lipowski}}]{Drofeli}%
  \BibitemOpen
  \bibfield  {author} {\bibinfo {author} {\bibfnamefont {M.}~\bibnamefont
  {Droz}}, \bibinfo {author} {\bibfnamefont {A.~L.}\ \bibnamefont {Ferreira}},
  \ and\ \bibinfo {author} {\bibfnamefont {A.}~\bibnamefont {Lipowski}},\
  }\href@noop {} {\bibfield  {journal} {\bibinfo  {journal} {Phys. Rev. E}\
  }\textbf {\bibinfo {volume} {67}},\ \bibinfo {pages} {056108} (\bibinfo
  {year} {2003})}\BibitemShut {NoStop}%
\bibitem [{\citenamefont {Marro}\ and\ \citenamefont
  {Dickman}(1999)}]{marro1999npt}%
  \BibitemOpen
  \bibfield  {author} {\bibinfo {author} {\bibfnamefont {J.}~\bibnamefont
  {Marro}}\ and\ \bibinfo {author} {\bibfnamefont {R.}~\bibnamefont
  {Dickman}},\ }\href@noop {} {\emph {\bibinfo {title} {{Nonequilibrium Phase
  Transitions in Lattice Models}}}}\ (\bibinfo  {publisher} {Cambridge
  University Press},\ \bibinfo {address} {Cambridge},\ \bibinfo {year}
  {1999})\BibitemShut {NoStop}%
\bibitem [{\citenamefont {Castellano}\ \emph
  {et~al.}(2009{\natexlab{b}})\citenamefont {Castellano}, \citenamefont
  {Mu\~noz},\ and\ \citenamefont {Pastor-Satorras}}]{PhysRevE.80.041129}%
  \BibitemOpen
  \bibfield  {author} {\bibinfo {author} {\bibfnamefont {C.}~\bibnamefont
  {Castellano}}, \bibinfo {author} {\bibfnamefont {M.~A.}\ \bibnamefont
  {Mu\~noz}}, \ and\ \bibinfo {author} {\bibfnamefont {R.}~\bibnamefont
  {Pastor-Satorras}},\ }\href@noop {} {\bibfield  {journal} {\bibinfo
  {journal} {Phys. Rev. E}\ }\textbf {\bibinfo {volume} {80}},\ \bibinfo
  {pages} {041129} (\bibinfo {year} {2009}{\natexlab{b}})}\BibitemShut
  {NoStop}%
\bibitem [{\citenamefont {Kaya}\ \emph {et~al.}(2000)\citenamefont {Kaya},
  \citenamefont {Kabak{\c c}io{\v g}lu},\ and\ \citenamefont
  {Erzan}}]{KKEmodel00}%
  \BibitemOpen
  \bibfield  {author} {\bibinfo {author} {\bibfnamefont {H.}~\bibnamefont
  {Kaya}}, \bibinfo {author} {\bibfnamefont {A.}~\bibnamefont {Kabak{\c c}io{\v
  g}lu}}, \ and\ \bibinfo {author} {\bibfnamefont {A.}~\bibnamefont {Erzan}},\
  }\href@noop {} {\bibfield  {journal} {\bibinfo  {journal} {Phys. Rev. E}\
  }\textbf {\bibinfo {volume} {61}},\ \bibinfo {pages} {1102} (\bibinfo {year}
  {2000})}\BibitemShut {NoStop}%
\bibitem [{\citenamefont {Corberi}\ \emph {et~al.}(2008)\citenamefont
  {Corberi}, \citenamefont {Lippiello},\ and\ \citenamefont
  {Zannetti}}]{PhysRevE.78.011109}%
  \BibitemOpen
  \bibfield  {author} {\bibinfo {author} {\bibfnamefont {F.}~\bibnamefont
  {Corberi}}, \bibinfo {author} {\bibfnamefont {E.}~\bibnamefont {Lippiello}},
  \ and\ \bibinfo {author} {\bibfnamefont {M.}~\bibnamefont {Zannetti}},\
  }\href@noop {} {\bibfield  {journal} {\bibinfo  {journal} {Phys. Rev. E}\
  }\textbf {\bibinfo {volume} {78}},\ \bibinfo {pages} {011109} (\bibinfo
  {year} {2008})}\BibitemShut {NoStop}%
\bibitem [{\citenamefont {Gardiner}(2010)}]{gardiner4ed2010}%
  \BibitemOpen
  \bibfield  {author} {\bibinfo {author} {\bibfnamefont {C.}~\bibnamefont
  {Gardiner}},\ }\href@noop {} {\emph {\bibinfo {title} {Stochastic Methods: A
  Handbook for the Natural and Social Sciences}}},\ \bibinfo {edition} {4th}\
  ed.\ (\bibinfo  {publisher} {Springer-Verlag},\ \bibinfo {address} {Berlin},\
  \bibinfo {year} {2010})\BibitemShut {NoStop}%
\bibitem [{\citenamefont {Castellano}\ and\ \citenamefont
  {Pastor-Satorras}(2011)}]{Castellano11}%
  \BibitemOpen
  \bibfield  {author} {\bibinfo {author} {\bibfnamefont {C.}~\bibnamefont
  {Castellano}}\ and\ \bibinfo {author} {\bibfnamefont {R.}~\bibnamefont
  {Pastor-Satorras}},\ }\href {\doibase 10.1103/PhysRevE.83.016113} {\bibfield
  {journal} {\bibinfo  {journal} {Phys. Rev. E}\ }\textbf {\bibinfo {volume}
  {83}},\ \bibinfo {pages} {016113} (\bibinfo {year} {2011})}\BibitemShut
  {NoStop}%
\bibitem [{\citenamefont {Sood}\ and\ \citenamefont
  {Redner}(2005)}]{PhysRevLett.94.178701}%
  \BibitemOpen
  \bibfield  {author} {\bibinfo {author} {\bibfnamefont {V.}~\bibnamefont
  {Sood}}\ and\ \bibinfo {author} {\bibfnamefont {S.}~\bibnamefont {Redner}},\
  }\href {\doibase 10.1103/PhysRevLett.94.178701} {\bibfield  {journal}
  {\bibinfo  {journal} {Phys. Rev. Lett.}\ }\textbf {\bibinfo {volume} {94}},\
  \bibinfo {pages} {178701} (\bibinfo {year} {2005})}\BibitemShut {NoStop}%
\bibitem [{\citenamefont {Masuda}\ \emph {et~al.}(2010)\citenamefont {Masuda},
  \citenamefont {Gibert},\ and\ \citenamefont {Redner}}]{PhysRevE.82.010103}%
  \BibitemOpen
  \bibfield  {author} {\bibinfo {author} {\bibfnamefont {N.}~\bibnamefont
  {Masuda}}, \bibinfo {author} {\bibfnamefont {N.}~\bibnamefont {Gibert}}, \
  and\ \bibinfo {author} {\bibfnamefont {S.}~\bibnamefont {Redner}},\ }\href
  {\doibase 10.1103/PhysRevE.82.010103} {\bibfield  {journal} {\bibinfo
  {journal} {Phys. Rev. E}\ }\textbf {\bibinfo {volume} {82}},\ \bibinfo
  {pages} {010103} (\bibinfo {year} {2010})}\BibitemShut {NoStop}%
\bibitem [{\citenamefont {Lambiotte}\ and\ \citenamefont
  {Redner}(2008)}]{Lambiotte08}%
  \BibitemOpen
  \bibfield  {author} {\bibinfo {author} {\bibfnamefont {R.}~\bibnamefont
  {Lambiotte}}\ and\ \bibinfo {author} {\bibfnamefont {S.}~\bibnamefont
  {Redner}},\ }\href@noop {} {\bibfield  {journal} {\bibinfo  {journal}
  {Europhys. Lett.}\ }\textbf {\bibinfo {volume} {82}},\ \bibinfo {pages}
  {18007} (\bibinfo {year} {2008})}\BibitemShut {NoStop}%
\bibitem [{\citenamefont {Slanina}\ \emph {et~al.}(2008)\citenamefont
  {Slanina}, \citenamefont {Sznajd-Weron},\ and\ \citenamefont
  {Przyby{\l}a}}]{Slanina08}%
  \BibitemOpen
  \bibfield  {author} {\bibinfo {author} {\bibfnamefont {F.}~\bibnamefont
  {Slanina}}, \bibinfo {author} {\bibfnamefont {K.}~\bibnamefont
  {Sznajd-Weron}}, \ and\ \bibinfo {author} {\bibfnamefont {P.}~\bibnamefont
  {Przyby{\l}a}},\ }\href {http://stacks.iop.org/0295-5075/82/i=1/a=18006}
  {\bibfield  {journal} {\bibinfo  {journal} {Europhys. Lett.}\ }\textbf
  {\bibinfo {volume} {82}},\ \bibinfo {pages} {18006} (\bibinfo {year}
  {2008})}\BibitemShut {NoStop}%
\end{thebibliography}

%

\end{document}